\documentclass[11pt,a4paper]{article}
\usepackage[T1]{fontenc}
\usepackage[utf8]{inputenc}
\usepackage[english]{babel}
\usepackage{amsmath, amssymb, amsthm}
\usepackage{graphicx}
\usepackage{flafter}
\usepackage{geometry}

\author{Yuri Yu. Tarasevich}
\title{Electrical conductivity of a two-dimensional two-component medium: discretization and effective medium approximation}
\date{Astrakhan Tatishchev State University,\\ 
Astrakhan, Russia, 414056, 20a Tatishchev St.\\
Tel.: (+7 8512) 246496 ext. 267, e-mail: tarasevich@asu-edu.ru
}
\begin{document}
\maketitle

\begin{figure}[!htb]
  \centering
  \includegraphics[width=\textwidth]{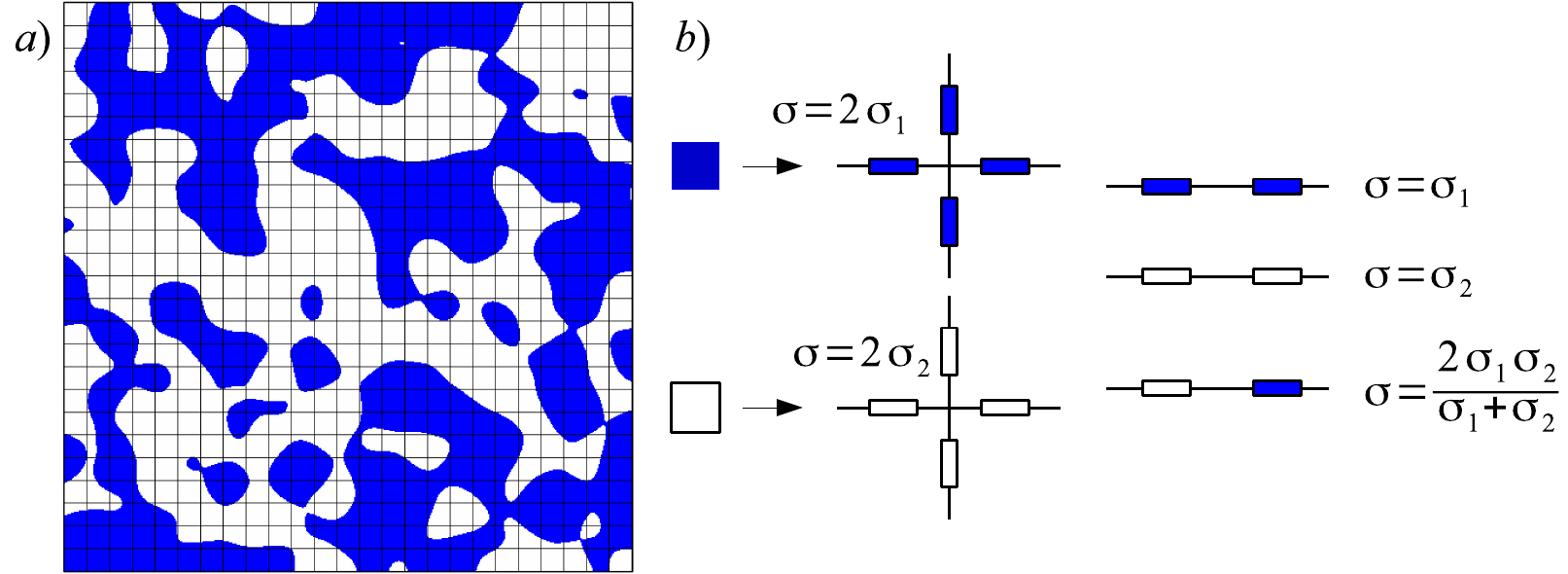}
\caption{a) Random two-dimensional two-component medium with an imposed grid. b) Transformation of cells into resistors.}\label{fig:transform}
\end{figure}

Let us consider a two-dimensional two-component medium in which the components have electrical conductivities $\sigma_1$ and $\sigma_2$, with $\sigma_1 \leqslant \sigma_2$. Let the second component occupy an area fraction $p$, and the first occupy a fraction $1-p$ (Fig.~\ref{fig:transform}a). If the components are distributed randomly and isotropically, and $p = 1/2$, then the effective conductivity of the medium is ${{\sigma }_{\text{eff}}}=\sqrt{{{\sigma }_{1}}{{\sigma }_{2}}}$~\cite{Dykhne1970}. Now we superimpose a square grid on the two-dimensional medium. We assume that a cell has conductivity $\sigma_1$ if most of the cell is occupied by the first component, and $\sigma_2$ otherwise. With each cell we associate a star of four resistors such that the junction point (node) corresponds to the cell center, and the free ends of the resistors correspond to the midpoints of the cell sides (Fig.~\ref{fig:transform}b). The set of all resistor stars forms a regular network. We assume that the resistors have conductances $2\sigma_1$ and $2\sigma_2$ depending on the cell in which they are located. Two series-connected resistors connecting the centers of neighboring cells are replaced by one equivalent resistor. If two neighboring cells both have conductivity $\sigma_1$, the equivalent resistor has conductance $\sigma_1$; if both have $\sigma_2$, then $\sigma_2$; if one cell has $\sigma_1$ and the other $\sigma_2$, the conductance of the equivalent resistor is
$$
\sigma_{12} = \frac{2\sigma_1 \sigma_2}{\sigma_1 + \sigma_2}.
$$

Thus, the effective network consists of resistors of three types: with conductance $\sigma_2$ (probability $p^2$), with conductance $\sigma_1$ (probability $(1-p)^2$), and with conductance $\sigma_{12}$ (probability $2p(1-p)$). This approach is analogous to~\cite{Grinchuk2026}.

Applying Foster's theorem~\cite{Foster1961} to this random regular resistor network using the approach of~\cite{Marchant1979} or the effective medium approximation in the form of~\cite{Kirkpatrick1973}, we find that the effective conductivity $\sigma$ of the network satisfies the equation
$$
\sum_{i} p_i \, \frac{\sigma_i - \sigma}{\sigma_i + (d/2 - 1)\, \sigma} = 0,
$$
where $p_i$ are the probabilities of the corresponding conductivities $\sigma_i$, and $d$ is the coordination number of the network (for a square network $d = 4$).

Substituting the probabilities and conductivities, we obtain
\begin{equation}\label{eq:EMA_general}
  (1-p)^2 \frac{\sigma_1 - \sigma}{\sigma_1 + \sigma}  + p^2  \frac{\sigma_2 - \sigma}{\sigma_2 + \sigma} + 2p(1-p) \frac{\sigma_{12} - \sigma}{\sigma_{12} + \sigma} = 0.
\end{equation}

Let us introduce dimensionless variables $x=\frac{\sigma }{\sqrt{{{\sigma }_{1}}{{\sigma }_{2}}}}$ and $r=\sqrt{\frac{{{\sigma }_{1}}}{{{\sigma }_{2}}}}$.
Then $\sigma_1 = r\sqrt{\sigma_1\sigma_2}$, $\sigma_2 = \frac{1}{r}\sqrt{\sigma_1\sigma_2}$, and \[\sigma_{12} = \frac{2r}{r^2+1}\,\sqrt{\sigma_1\sigma_2}.\]
Note that $\Delta = \frac{\sigma_2}{\sigma_1}$ is the electrical contrast, so $r = \sqrt{\Delta^{-1}}$.
Using the dimensionless variables, equation~\eqref{eq:EMA_general} takes the form
\begin{equation}\label{eq:eq2}
(1-p)^2 \frac{r-x}{r+x} + p^2 \frac{1-r x}{ 1 + r x} + 2p(1-p)  \frac{2r - x (r^2+1)}{ 2r + x (r^2+1)} = 0.
\end{equation}
Note that $x(0)=r$ and $x(1)={{r}^{-1}}$. If $r=0$, then~\eqref{eq:eq2} simplifies to $-(1-p)^2  + p^2  - 2p(1-p) = 0,$ which has only one positive root $p_c = 2^{-1/2}$. As $r\to 0$, the bonds with conductivity $\sigma_{12}$ also vanish, the fraction of conducting bonds is $p^2$, and $p_c^2 =1/2$ coincides with the duality value for a two-dimensional two-component system. Therefore, $p_c$ can be identified with the critical concentration.

\begin{figure}[!htb]
  \centering
  \includegraphics[width=0.45\textwidth]{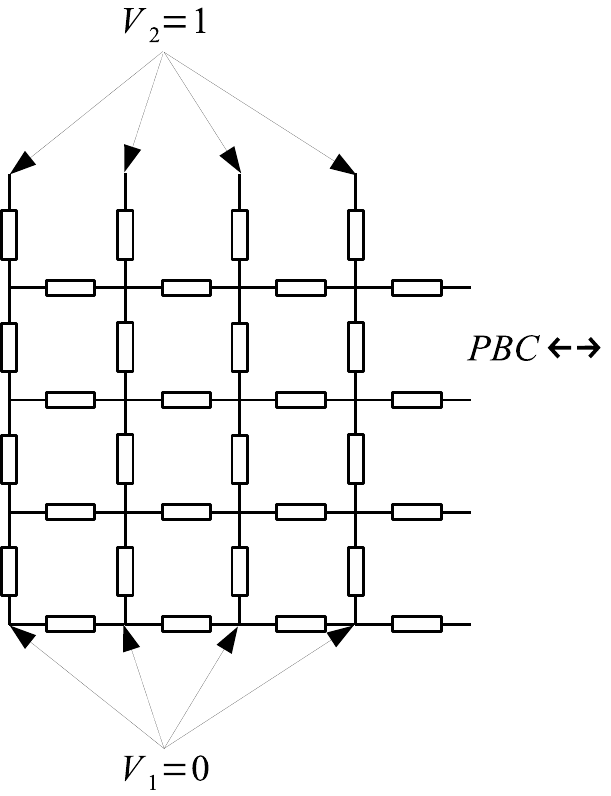}
  \caption{Resistor network}\label{fig:lattice}
\end{figure}

After mathematical transformations, we obtain a cubic equation of the form
\begin{equation}\label{eq:EMAgeneralpoly}
x^3 - C_2 x^2 - C_1 x - \frac{2r}{r^2+1} = 0,
\end{equation}
where \[C_1 = \frac{(3 - 4p) r^2 + (4p - 1)}{r^2 + 1}\] and \[C_2 = (1 - 4p + 2p^2) r + (2p^2 - 1) \frac{1}{r} - (1 - 2p)^2 \frac{2r}{r^2+1}.\]
These expressions are symmetric under the substitution $p \mapsto 1-p$, $r \mapsto 1/r$, which corresponds to the interchange of the components $\sigma_1 \leftrightarrow \sigma_2$.
At $p = 1/2$, the coefficients reduce to \[C_1 = 1, \qquad C_2 = -\frac{1}{2}\left( r + \frac{1}{r} \right),\] and the cubic equation takes the form \[x^3 + \frac{1}{2}\left( r + \frac{1}{r} \right)x^2 - x - \frac{2r}{r^2+1} = 0.\]
For the calculations, a square resistor network of size \(L \times L\) nodes was used (Fig.~\ref{fig:lattice}). The nodes are labeled by indices \((i, j)\), where \(i = 1, \ldots, L\) is the row number (from top to bottom), and \(j = 1, \ldots, L\) is the column number (from left to right). On the top and bottom boundaries, we impose Dirichlet conditions: the potentials of all nodes in the top row \(i = 1\) are fixed and equal to \(V = 1\), and the potentials of all nodes in the bottom row \(i = L\) are equal to \(V = 0\). On the side boundaries, we impose periodic boundary conditions: node \((i, L)\) is connected to node \((i, 1)\) by a horizontal resistor for each row \(i = 1, \ldots, L\). Thus, the network is topologically equivalent to a cylinder closed in the horizontal direction and open in the vertical direction. The boundary conditions are imposed directly on the nodes, and all resistors of the network are treated on an equal footing.

The conductances of the resistors take three possible values:
\[
\sigma_1 = r^2, \qquad \sigma_2 = 1, \qquad \sigma_{12} = \frac{2}{1 + r^{-2}},
\]
where \(0 < r \leqslant 1\) is the contrast parameter. In the case of uncorrelated disorder, each resistor is independently assigned one of these values with probabilities
\[
P(\sigma_1) = (1-p)^2, \qquad P(\sigma_2) = p^2, \qquad P(\sigma_{12}) = 2p(1-p),
\]
where \(p \in [0, 1]\). In the case of correlated disorder, each node is independently labeled ``white'' with probability \(p\) or ``black'' with probability \(1-p\), and the conductance of a resistor is determined by the colors of the two nodes it connects:
\[
\sigma = \begin{cases}
1, & \text{white--white}, \\[4pt]
r^2, & \text{black--black}, \\[4pt]
\dfrac{2}{1 + r^{-2}}, & \text{white--black}.
\end{cases}
\]
In both cases, the marginal probabilities of the three values coincide, but the correlated model generates spatial correlations between the conductances of neighboring resistors.

The effective conductance of the network \(G_{\text{eq}}\) is defined as the total current flowing into the top row of nodes when the potential difference between the top and bottom rows is unity. A network of size \(L=200\) was used. The results were averaged over 100 independent realizations.
Fig. 3 presents the results of numerical solution of equation \eqref{eq:EMAgeneralpoly} for several values of $p$ at fixed $r$.
\begin{figure}[htb]
  \centering
  \includegraphics[width=0.55\textwidth]{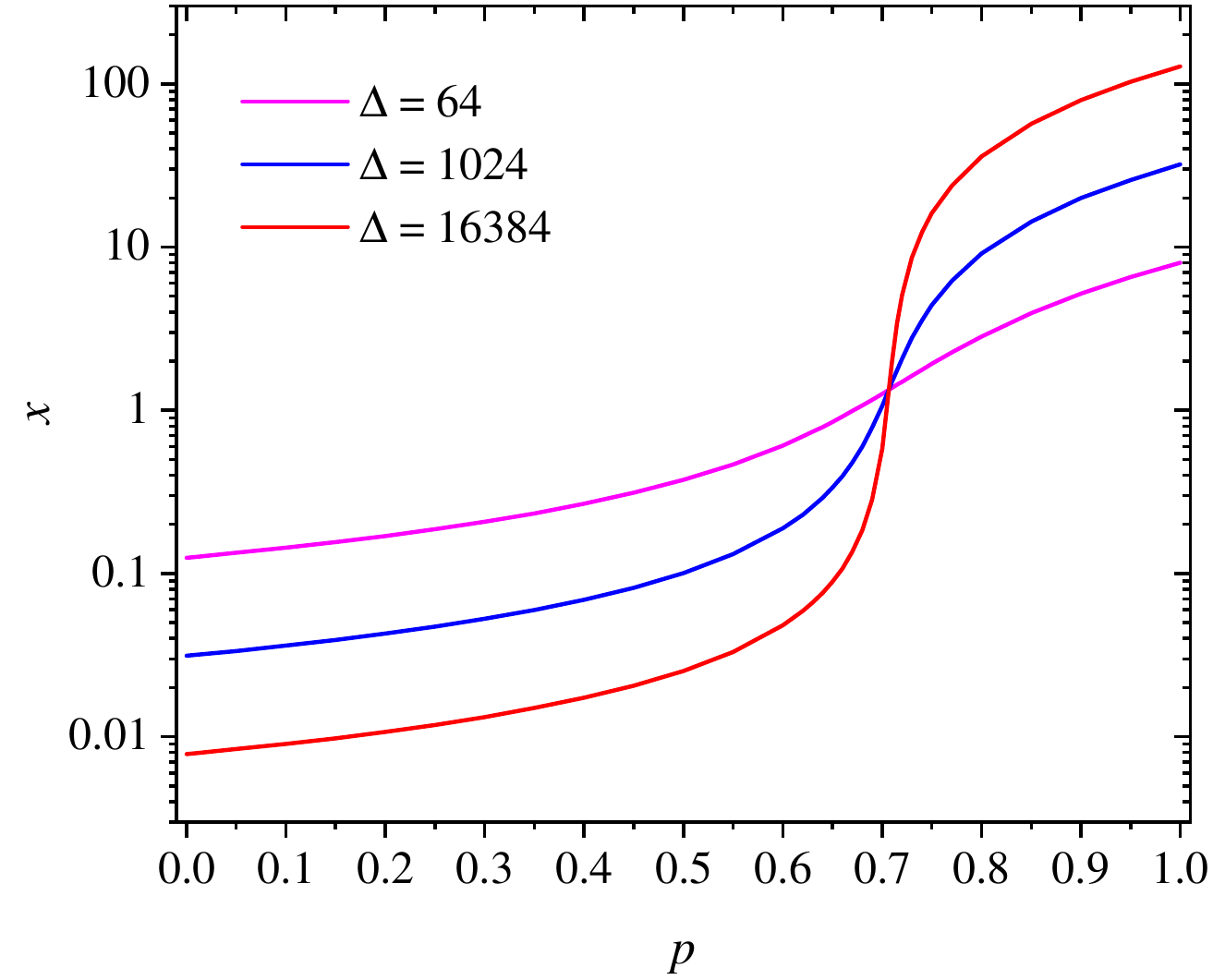}
  \caption{Dependence of the dimensionless effective conductivity $x$ on the fraction $p$ of the conducting component for various values of the conductivity ratio.}\label{fig:x-vs-p}
\end{figure}

\begin{figure}[htb]
  \centering
  \includegraphics[width=0.55\textwidth]{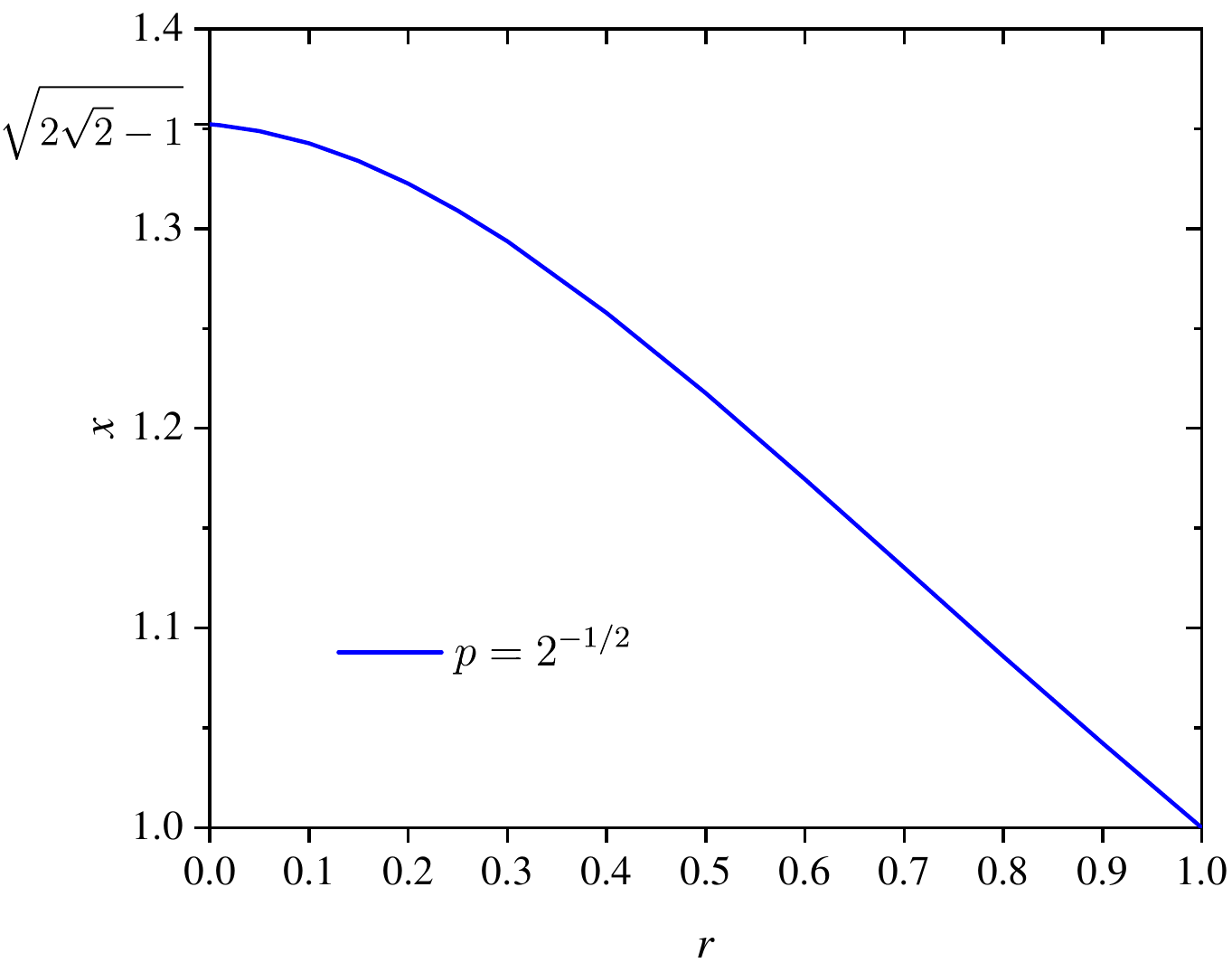}
  \caption{Dependence of the dimensionless effective conductivity $x$ on the conductivity ratio $r$ at \(p=1/\sqrt{2}\).}\label{fig:x-vs-r}
\end{figure}

Although the curves in Fig.~\ref{fig:x-vs-p} appear to intersect at one point, this apparent universality is absent. Fig.~\ref{fig:x-vs-r} presents the results of numerical solution of equation~\eqref{eq:EMAgeneralpoly} at $p=p_c$ for several values of $r$: the value of $x$ at $p=p_c$ still depends on $r$, so the curves do not actually meet at one point.

\begin{figure}[htb]
  \centering
  \includegraphics[width=0.55\textwidth]{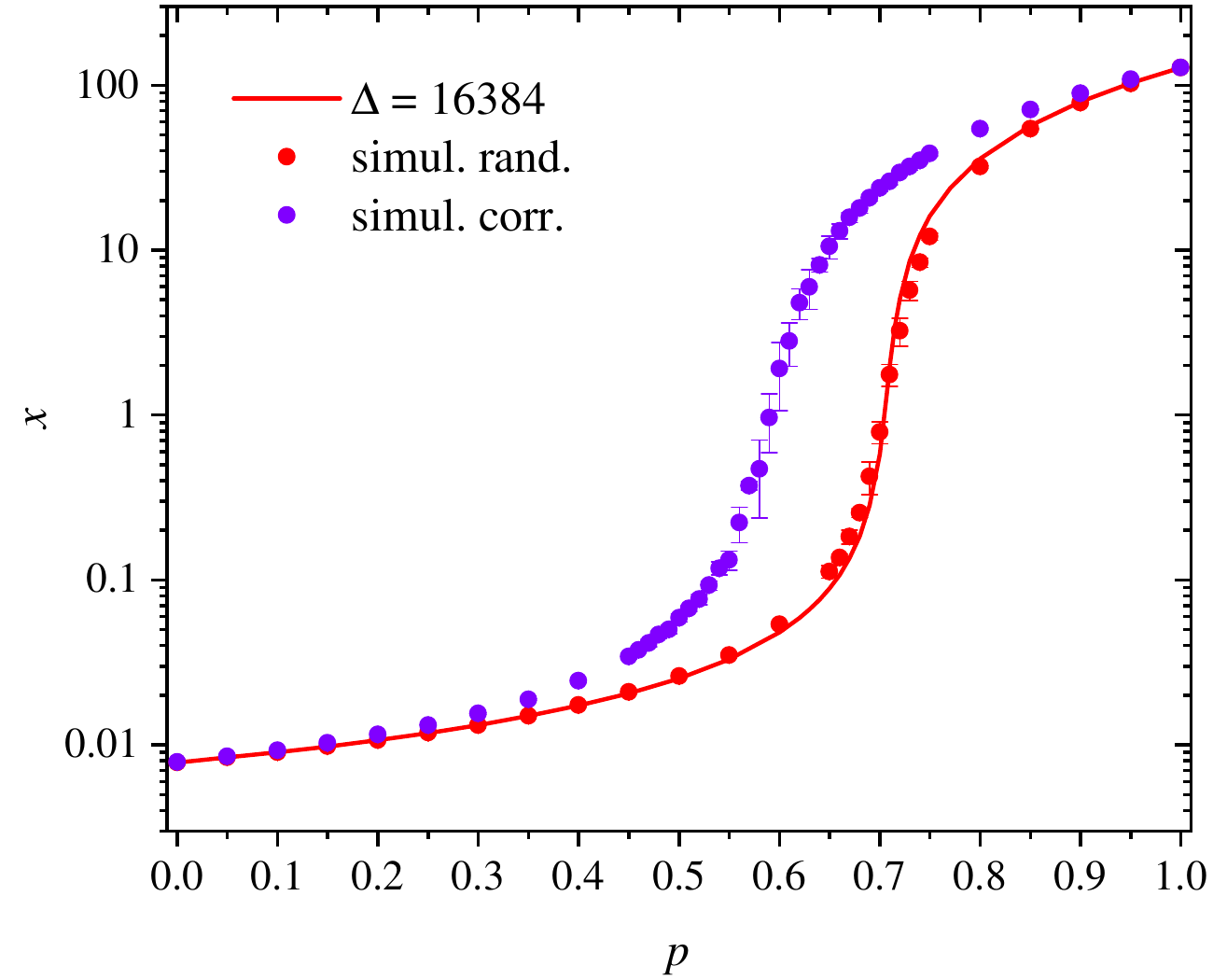}
  \caption{Dependence of the dimensionless effective conductivity $x$ on the fraction $p$ of the conducting component at \(r=2^{-7}\). The lines correspond to the effective medium approximation, and the markers to the results of computer simulation.}\label{fig:corr-vs-rnd}
\end{figure}

Fig.~\ref{fig:corr-vs-rnd} shows the dependence of the effective electrical conductivity on the concentration of the conducting component at a fixed contrast value. The line represents the predictions of the effective medium theory. The data points correspond to direct calculations performed on a \(200 \times 200\) resistor network, averaged over ten independent systems. When resistors of three different types are distributed randomly throughout the system, good agreement is observed between the predictions of the effective medium theory and direct calculations. In contrast, when the resistance values are correlated---following the discretization rules applied to the original system (Fig.~\ref{fig:transform})---significant discrepancies arise between direct calculations and theoretical predictions.

The inflection point of the graph, which corresponds to the percolation threshold in the case of infinite conductivity contrast, is strongly shifted to the right. For a two-dimensional two-component insulator--conductor system, the phase transition point must be exactly equal to $1/2$. The shift of the critical concentration to the right is a consequence of correlations. The resulting resistor network is strongly correlated: the four resistors emanating from a single node are guaranteed to have identical resistances. Thus, comparison of the predictions of the effective medium approximation with direct calculations casts doubt on the approach proposed in~\cite{Grinchuk2026}. In addition, the significant discrepancy between the predictions of the effective medium approximation and direct calculations for random nanowire networks~\cite{Tarasevich2025} becomes clear.

\bibliographystyle{plain}
\bibliography{{../EMT2026short}}

\end{document}